\documentclass[epj]{svjour}
\usepackage{graphics}
\usepackage{epstopdf}
\usepackage{amsmath}
\usepackage{amssymb}
\begin{document}
\title{Kerr/CFT Correspondence on Kerr-Newman-NUT-Quintessence Black Hole}
\author{Muhammad F. A. R. Sakti\inst{1}\thanks{\emph{e-mail:} m.fitrah@students.itb.ac.id} \and Agus Suroso\inst{1,2}\thanks{\emph{e-mail:} agussuroso@fi.itb.ac.id} \and Freddy P. Zen\inst{1,2}
\thanks{\emph{e-mail:} fpzen@fi.itb.ac.id}%
}                     
%
%
\institute{Theoretical Physics Laboratory, THEPI Division, Institut Teknologi Bandung, Jl. Ganesha 10 Bandung, 40132, Indonesia \and Indonesia Center for Theoretical and Mathematical Physics (ICTMP), Institut Teknologi Bandung, Jl. Ganesha 10 Bandung, 40132, Indonesia}
\date{Received: date / Revised version: date}
%
\abstract{
Rotating black hole solution surrounded by quintessential matter is recently discussed because it might be the promising solution to study the effect of dark energy in small scale of the universe. This quintessential solution is originally derived from the condition of additivity and linearity for the energy-momentum tensor. We carry out the thermodynamic properties of this solution using the Kerr/CFT correspondence for several specific quintessential equation of state parameters.  A problem arises when we compute the central charge because the canonical conserved charge is needed to be calculated from the Lagrangian. However, the exact Lagrangian of the quintessence is not defined yet in the original derivation. Yet we solve this problem by the assumption that there is only a contribution from gravitational field to the central charge. Then we could find the entropy of this black hole after calculating the temperature and using Cardy entropy formula. Another problem comes out when the spin goes to zero to find the Reissner-Nordstr{\"o}m-NUT-Quintessence solution. To solve it, we extend to the 5-dimensional solution. In the end, we obtain the entropy for this 5-dimensional solution. So the quintessential black hole solution is dual with the CFT.
} 
\maketitle
\section{Introduction}
\label{sec:intro}

Black holes become local the manifestations of space-time's curvature which correspond to Einstein's theory of gravity. On a larger scale, Einstein's theory of gravity can also predict the formation of the universe until the present condition. It has been proved that our universe in the present era is experiencing accelerated expansion due to the existence of a unique matter known as dark energy \cite{Copeland2006}. Dark energy provides the largest contribution to our universe besides the visible matter and dark matter. Numerous models have been proposed to explain the role of dark energy in the universe which of course must generate the fact that the universe is accelerated and expanding. A model that is quite simple and can explain this phenomenon is quintessence \cite{Copeland2006}. This quintessence model is fine enough to describe dark energy even though many other models are proposed to be able to produce data that corresponds to an accelerated expanding universe such as in \cite{Arianto2007,Arianto2008,Zen2009,Arianto2010,Feranie2010,Arianto2011,Suroso2013,Suroso2015,Getbogi2016}.  

If indeed this model is suitable for dark energy, the existence of the quintessence could affect a local gravitational field in the universe. For instance, black holes can experience interaction with the quintessence. In \cite{Kiselev2003}, it is formulated how the existence of the quintessence affects the solution of the black holes by performing the condition of additivity and linearity for the energy-momentum tensor. It is because the Lagrangian of the scalar field is not defined in that derivation. There are parameters such the equation of state (EoS) parameter and density related to the quintessence in the black hole solution. When the quintessence is considered as the cause of the accelerated expanding universe, the EoS parameter is bounded in the range $ -1 <\omega <-1/3 $. However, when $ -1/3 <\omega <0 $, the black hole solution possesses an asymptotically flat solution. The rotating counterpart of this black hole solution is calculated employing the Newman-Janis algorithm in \cite{Toshmatov2017,Wang2017} (originally shown in \cite{Newman1965,Newmanetal1965}). It is also extended to the solution with the presence of NUT charge in Rastall gravity in \cite{Sakti2019} by employing Demia{\'n}ski-Newman-Janis algorithm \cite{Erbin2015,Erbin2015a,Erbin2016,Erbin2017}. The other extension of this solution with some additional constants and in modified gravity theories can be found in \cite{Kumar2018,Wang2018,Maharaj2017,Lee2018}.

Because of the quintessence's existence on the black hole solutions, it is fascinating to learn more about black hole properties especially which are related to the quantum aspect. Macroscopic entropy of this black hole has generally been derived in \cite{Sakti2019} with the presence of the Rastall parameter. However, it is more interesting to learn from a microscopic point of view. In addition to calculating the microstate, as in the black hole \cite{Vafa1996}, we can make use of the Kerr/CFT correspondence to compute the microscopic entropy where it has been successful to calculate the entropy for some black hole solutions \cite{Guica2009,Hartman2009,Ghezelbash2009,Lu2009,Li2010,Anninos2010,Ghodsi2010,Ghezelbash2012,Astorino2015,Astorino2015a,Siahaan2016,Astorino2016,Sinamuli2016,Compere2017,Sakti2018,Sakti2019b}.  Yet a problem arises because there is no exact Lagrangian form of the the original solution of black holes surrounded by the quintessence. Because of that, we will try to figure out the way to solve this problem.

In this letter, we work on the Kerr/CFT correspondence for the special case of quintessential black hole found in \cite{Sakti2019} with vanishing Rastall parameter and by taking certain EoS parameter which are $ \omega = 0, 1/3, $ and $-1/3 $. These EoS values denote the dust-dominated type, radiation-dominated type and in the lower limit of equation of state for accelerated expansion universe, respectively. Then it is found that the near-horizon form of this space-time metric resembles the usual NHEK black hole \cite{Guica2009,Compere2017,Sakti2016,Sakti2019a}. As we mentioned before, the problem arises because we do not have the exact form of the Lagrangian. Yet commonly the simplest model of the quintessence is the real scalar field entering
the action with a canonical kinetic term and an exponential
potential. To solve it, we assume that the scalar field has no contribution to the central charge and it is proved in \cite{Compere2017} for general scalar fields. In addition to that, the electromagnetic contribution is also neglected. 

Regarding the non-gravitational fields have no contribution in the central term, the central charge is found and contains the quintessential density parameter. The effect of this density is different in every type of quintessential black hole solution that we consider. To complete the calculation, we also derive the conformal temperature by assuming the quantum scalar field around the extremal black hole solutions. As the Kerr/CFT correspondence \cite{Guica2009}, we apply the famous Cardy entropy to calculate the entropy from CFT. Finally we find that it agrees with the Bekenstein-Hawking entropy. Then we also extend the 4-dimensional solution to the higher dimensional solution to study the Reissner-Nordstr{\"o}m-NUT-Quintessence black holes solution. We apply the similar correspondence to calculate the microscopic entropy from CFT. In the end, it is found that the entropy of 5-dimensional solution is still in agreement with the Bekenstein-Hawking one. So it is clear that the quintessential black hole solution could be dual with CFT even it has no exact Lagrangian form of the quintessence. 

The remaining parts of this paper are structured as follows.
In section 2, we present the Kerr-Newman-NUT-Quintessence black hole solution with general value of EoS parameter and its several macroscopically thermodynamic properties. In the next section, the near-horizon form of the metric and electromagnetic potential are derived using the appropriate coordinates transformation. After finding the near-horizon form, we calculate the central charge applying the asymptotic symmetry group and the CFT temperature based on the Frolov-Thorne vacuum in section 4. Later on, the extended 5-dimensional solution, Reissner-Nordstr{\"o}m-NUT-Quintessence black hole, is derived. We also employ the Kerr/CFT correspondence to this solution to compute the microscopic entropy. In the last section, we give the summary of the results of this paper.

\section{KNUTQ black hole solution}
\label{sec:KNUTK}
The original solution of black hole surrounded by the quintessence is found by Kiselev \cite{Kiselev2003}. Then the rotating counterpart is derived in \cite{Toshmatov2017,Wang2017,Sakti2019,Wang2018,Kumar2018,Maharaj2017,Lee2018}. Nevertheless, we want to focus on the general solution that has been derived in \cite{Sakti2019} with the existence of NUT charge and with additional magnetic charge. Accordingly, we will carry out the more general one which is the Kerr-Newman-NUT-Quintessence black hole solution. Its thermodynamic properties are going to be shown in this section. This black hole solution has been derived by employing the Demia{\'n}ski-Newman-Janis algorithm to the dyonic Reissner-Nordstr{\"o}m-Quintessence black hole solution within the Rastall's theory of gravity. Within this modified gravity theory, the non-conservative energy-momentum tensor is considered. Yet herein the paper, the Rastall's parameter will be turned off thereupon it reduces to the solution of Einstein's theory of gravity. The space-time metric is given by
\begin{eqnarray}
ds^2 &=& -\frac{\Delta}{\rho^2}\left[dt - \left( a\text{sin}^2\theta +2n(1-\text{cos}\theta) \right) d\varphi \right]^2+ \frac{\rho^2}{\Delta}dr^2 \nonumber\\
& &+\rho ^2 d\theta ^2 +\frac{\text{sin}^2\theta}{\rho ^2}\left[adt-(r^2+(a+n)^2) d\varphi \right]^2 , \label{eq:metricresult}
\end{eqnarray}
where
\begin{equation}
\Delta = r^2 - 2Mr+a^2+e^2+g^2-n^2 -\alpha r^{1-3\omega} , \nonumber\
\end{equation}
\begin{equation}
\rho^2 = r^2 + (n+a\text{cos}\theta)^2. \label{eq:delta}\
\end{equation}

Respectively, the parameters on metric (\ref{eq:metricresult}) are mass $ M $, spin $ a $, electric charge $ e $, magnetic charge $ g $, NUT charge $ n $, density of the quintessence $ \alpha $ and the equation of state parameter of the quintessence $ \omega $. For the quintessential matter, $ \omega $ has value in range $ -1 < \omega < 0 $ where $ -1 < \omega < -1/3 $ is indicating the accelerating expansion and $ -1/3 < \omega < 0 $ makes the asymptotically flat solution. Afterwards, we define $ e^2 +g ^2 = q^2 $ for simplicity. The space-time metric (\ref{eq:metricresult}) is derived in \cite{Sakti2019} by firstly deriving the Reisser-Norst{\"o}rm-Quintessence black hole in Rastall gravity. The quintesssential part is derived from the condition of additivity and linearity of energy-momentum tensor. To be specific, this principle reads $ Q^t_t = Q ^r_r $ where $ Q^\mu_\nu $ is the energy-momentum tensor of the quintessence. The general expression for the energy–momentum tensor of quintessence is given by
\begin{eqnarray}
&& Q^t_t = -\rho_q(r), \\
&& Q^i_j = 3\rho_q(r)\omega\left[(1+3B)\frac{r_ir^j}{r_nr^n}-B\delta_i^j \right],
\end{eqnarray}
where $ B $ is just an arbitrary parameter. After isotropic averaging over the angles, we obtain
\begin{eqnarray}
\langle T_i^j\rangle = \rho_q(r) \omega \delta_i^j  = \rho_q(r) \delta_i^j,
\end{eqnarray}
and we put $\langle r_ir^j \rangle = \frac{1}{3}\delta_i^j r_nr^n$. As the consequence, one could find
\begin{eqnarray}
Q^t_t = Q ^r_r = -\rho_q(r), ~~~
Q^2_2 = Q^3_3 = \frac{1}{2}(3\omega+1)\rho_q(r). \label{eq:quinenermom} \
\end{eqnarray}
After that, to add rotation and twisting parameter, they employ Demia{\'n}sky-Newman-Janis algoritm. The space-time metric (\ref{eq:metricresult}) possesses a number of horizons that depends on the quintessential EoS parameter. The electromagnetic potential related to the metric (\ref{eq:metricresult}) is given by \cite{Sakti2019}
\begin{eqnarray}
A_\mu dx^\mu &=& \frac{-er\left[a dt- \left\{ a^2\text{sin}^2\theta +2an(1-\text{cos}\theta) \right\} d\varphi \right]}{a\rho ^2}  \nonumber\\
&-&\frac{g(n+a\text{cos}\theta)\left[a dt- \left\{ r^2 + (a+n)^2 \right\} d\varphi \right]}{a\rho^2} .\ \label{eq:electromagneticpotential}
\end{eqnarray}
This potential is similar with the electromagnetic potential of the Kerr-Newman-NUT black hole\cite{Sakti2018,Podolsky2009}. The presence of the quintessential matter does not modify the form of the electromagnetic potential.

It is common that for vanishing value of density of the quintessence, the space-time metric (\ref{eq:metricresult}) is  a solution of the Einstein-Maxwell system. When there exists the quintessence, it is not only the solution of the Einstein-Maxwell system anymore. In cosmology, the quintessence could be modeled by the scalar field that makes the accelerated expansion of the universe. In the original paper \cite{Kiselev2003}, the black hole solution with the presence of the quintessential matter is derived not from the action, yet from the condition of additivity and linearity for the energy-momentum tensor.

Some properties of the solution (\ref{eq:metricresult}) are portrayed in term of event horizon $ r_+ $. The Hawking temperature is given by \cite{Sakti2019}
\begin{eqnarray}
T_H = \frac{2(r_+ - M) - (1-3\omega)\alpha r_+^{-3\omega}}{4\pi [r_+^2 + (a+n)^2]}, \label{eq:Hawkingtemp}\ 
\end{eqnarray}
where the mass is identified as
\begin{eqnarray}
M = \frac{1}{2r_+}\left(r_+ ^2 - \alpha r_+^{1-3\omega} +a^2+q^2-n^2 \right). \label{eq:massdef}
\end{eqnarray}
The angular velocity and Coulomb electromagnetic potential are given by \cite{Sakti2019}
\begin{equation}
\Omega_H = \frac{a}{r_+^2 + (a+n)^2}, \label{eq:angmomen}\\
\end{equation}
\begin{eqnarray}
\Phi_H &=& \Phi_e +\Phi_g, \nonumber\\
&=&  \frac{er_++g(a+n)}{r_+^2+(a+n)^2} -\frac{g(a+n)}{r_+^2+(a+n)^2} .  \label{eq:coulombpotential} \
\end{eqnarray}
Due to the existence of the quintessence, we should treat the constant $ \alpha $ as a thermodynamic variable \cite{Chen2008,Sekiwa2006}. Hence, we have the generalized force
\begin{eqnarray}
\Theta_H =\frac{\partial M}{\partial \alpha}\bigg|_{r=r_+}=-\frac{1}{2}r_+^{-3\omega},
\end{eqnarray}
when $ S,J,Q_e,Q_g $ are held constant. The entropy of the Kerr-Newman-NUT-Quintessence black holes solution is given by the usual Bekenstein-
Hawking relation \cite{Sakti2019}
\begin{eqnarray}
S_{BH} = \frac{A_{BH}}{4} = \pi \left(r_+^2 + (a+n)^2 \right). \label{eq:bhentropy}
\end{eqnarray}

In the next section, we are going to study the conjectured Kerr/CFT correspondence for some special cases of matter domination that is characterized by $ \omega $. As we know in the Kerr/CFT correspondence, the Lagrangian plays an important role in the definition of the canonical conserved charge that  will lead to the Virasoro algebra. This Virasoro algebra will lead to the central term that results in the central charge. This central charge is very important to be applied in Cardy entropy formula. So in the next section, we are going to figure out whether the black hole solution from those principles is still sufficient to be used to investigate the Kerr/CFT correspondence. So we wish to match the entropy from the Bekenstein-Hawking entropy to the entropy from the CFT one by employing the Kerr/CFT correspondence.

We will study some special conditions of the quintessential equation of state. The dust domination (DD) and radiation domination (RD) types will be implemented because there are no cosmological horizon on these types. For DD type black hole solution, the equation of state is identified by $ \omega = 0 $. While for RD type black hole solution, this parameter has the value of $ \omega = 1/3 $. The assumption of any kind of matter domination is based on what happened in our universe in cosmological theory. In addition, we wish to know for the solution when $ \omega = -1/3 $ because this value is still in the possibility of accelerating expansion universe (AE). These are done also for the simplicity because within this assumption the horizon will remain to be inner and outer horizons only. Therefore there is no cosmological horizon.

\section{Near-horizon Geometry of KNUTQ Black Hole}
In this section, we wish to show the near-horizon form of the quintessential black hole space-time (\ref{eq:metricresult}) for DD, RD, and AE conditions in order to study the Kerr/CFT correspondence. Regarding the black hole solution (\ref{eq:metricresult}) with the chosen conditions of $ \omega $, the roots of $ \Delta $ in Eq. (\ref{eq:delta}) are \cite{Sakti2019}
\begin{equation}
r_\pm^{DD} = \left(M+\frac{\alpha}{2}\right)\pm \sqrt{\left(M+\frac{\alpha}{2}\right)^2 +n^2 - a^2 -q^2}, \
\end{equation}
\begin{equation}
r_\pm^{RD} = M \pm \sqrt{M^2+n^2 +\alpha-a^2-q^2} , \
\end{equation}
\begin{equation}
r_\pm^{AE} = \frac{M \pm \sqrt{M^2-(a^2+q^2-n^2)(1-\alpha)}}{1-\alpha}. \label{eq:roots} \
\end{equation}
The extremalities for DD, RD, and AE types, respectively are defined as
\begin{equation}
\left(M+\frac{\alpha}{2}\right)^2+ n^2 = a^2 +q^2, \
\end{equation}
\begin{equation}
M^2+n^2 +\alpha = a^2+q^2 , \
\end{equation}
\begin{equation}
M^2 = (a^2+q^2-n^2)(1-\alpha). \label{eq:extremality} \
\end{equation}
Since we wish to study the region very near to the horizon, it is needed to define some new coordinates. In these new coordinates, we will also apply the extremal condition for every condition of $ \omega $. In order to get the near-horizon form of the metric (\ref{eq:metricresult}), we need to define the following coordinates transformation based on \cite{Guica2009,Compere2017} which is given by
\begin{eqnarray}
r = r_+ + \epsilon r_0 y, ~ t = \frac{r_0}{\epsilon}\tau, \varphi = \phi +\Omega_H \frac{r_0}{\epsilon}\tau , \label{eq:coortransmatterdom}\
\end{eqnarray}
where the scale $ r_0 $ is introduced to factor out the overall scale of the near-horizon geometry. We define that scale as $ r_0^2 = r_+^2 +(a+n)^2 $. The constant $ \epsilon $ is the infinitesimal parameter. Near the horizon of the extremal black hole, the function $ \Delta $ takes form
\begin{eqnarray}
\Delta = (r- r_+)^2 V + \mathcal{O}\left((r- r_+)^3\right).
\end{eqnarray}
The values of the function $ V $ are $ 1 $ for $ \omega=0,1/3 $ and $ (1-\alpha) $ for $ \omega=-1/3 $. After taking $ \epsilon \rightarrow 0 $, the near-horizon extremal metric is then given by
\begin{eqnarray}
ds^2 &=&\frac{\rho_+^2}{V} \left(-V^2 y^2d\tau^2 + \frac{dy^2}{y^2}+ V d\theta ^2 \right) \nonumber\\
& &+ \frac{r_0^4 \text{sin}^2\theta}{\rho_+^2} \left( d\phi + \frac{2ar_+}{r_0^2} yd\tau \right)^2 ,\label{eq:RDnearhor2} \
\end{eqnarray}
where 
\begin{eqnarray}
\rho_+^2 = r_+^2 +(n+a \text{cos}\theta)^2. \nonumber\
\end{eqnarray}
We wish to throw away the constant $ V^2 $ in front of the $ dt^2 $. As we know this space-time metric is stationary, so we could perform scaling over $ d\tau $ with $ d\tau \rightarrow d\tau /V $. Hence, the resulting metric is then
\begin{eqnarray}
ds^2 &=&\frac{\rho_+^2}{V} \left(-y^2d\tau^2 + \frac{dy^2}{y^2}+ V d\theta ^2 \right) \nonumber\\
&& + \frac{r_0^4 \text{sin}^2\theta}{\rho_+^2} \left( d\phi + k yd\tau \right)^2 ,\label{eq:RDnearhor3}\
\end{eqnarray}
where $ k $ is given by
\begin{eqnarray}
&& k = \frac{2ar_+}{V r_0^2}.
\end{eqnarray}
Straightforwardly, we could perceive that the near-horizon metric (\ref{eq:RDnearhor3}) contains AdS structure on the coordinates $( \tau, y )$. It augurs that the metric is not asymptotically flat anymore.

The coordinates transformation (\ref{eq:coortransmatterdom}) also governs the change on the electromagnetic potential. In order to do so, the electromagnetic potential (\ref{eq:electromagneticpotential}) is supposed to be expanded in $ (r-r_+) = \epsilon r_0 y $. Later on when we set $ \epsilon \rightarrow 0$, we have to remove the singularity by employing the gauge transformation, so we arrive at
\begin{eqnarray}
A_\mu dx^\mu = f(\theta) \left(d\phi +k yd\tau \right)-\frac{e[r_+^2-(a+n)^2]}{[r_+^2+(a+n)^2]}d\phi ,\ \label{eq:nearhorizonelectromagneticpot1}
\end{eqnarray}
where
\begin{eqnarray}
f(\theta) &=& \frac{ar_0^2(2gr_+ -2en-ae\text{cos}\theta)\cos\theta }{2ar_+ \rho_+ ^2} \nonumber\\
&& + \frac{r_0^2\left[2gr_+n +e(r_+^2-n^2) \right]}{2ar_+ \rho_+ ^2}. \label{eq:ftheta} \
\end{eqnarray}
This is the near-horizon form of the vector field related to the DD, RD, and AE types of black hole solution. This form is similar with  the Kerr-Newman-NUT black hole when $ r_+ =M $ \cite{Sakti2019} because the quintessence does not affect the form of the electromagnetic field.

It is clearly seen that this form of metric (\ref{eq:RDnearhor2}) resembles the near-horizon metric shown in \cite{Hartman2009} for Kerr-Newman-AdS or for more general black hole solutions are reviewed in \cite{Compere2017}. Therefore the isometry that is generated by the following vector fields
\begin{eqnarray}
& &\zeta_0 = \partial_\phi, ~ X_1 = \partial_\tau, X_2 = \tau \partial_\tau - y \partial_y , \nonumber\\
& & X_3 = \left(\frac{1}{2y^2}+\frac{\tau^2}{2} \right)\partial_\tau -\tau y \partial_y - \frac{k}{y}\partial_\phi .\label{eq:isometrynearhorizon}
\end{eqnarray}
It is worth noting that all of these isometries act within a three dimensional slice of fixed polar angle $ \theta $. It is  concluded that metric (\ref{eq:RDnearhor2}) enhanced by the $ SL(2,R) \times U(1) $ isometry. This isometry is a hint to adopt the Brown and Henneaux approach \cite{Brown1986} to find the central charge of the holographic dual conformal field theory description of an extremal rotating quintessential black hole.

\section{Entropy from CFT of KNUTQ Black Hole}
In the previous section, we saw the gravitational side of the Kerr/CFT correspondence. Later in the following, we are supposed to explore the quantum field theories side where the Cardy formula will be governed to calculate the entropy from the CFT. The Brown-Henneaux approach \cite{Brown1986} is needed to find the central charge of the dual holographic conformal field theory description of the black hole solutions. As we mentioned before, we face the problem in computing the canonical conserved charges of the quintessential matter since this solution comes from the condition of additivity and linearity for the energy-momentum tensor \cite{Kiselev2003}, not from the specific Langrangian form. Yet commonly the simplest model of the quintessence is the real scalar field entering the action with a canonical kinetic term and an exponential potential
\begin{eqnarray}
\mathcal{L}_{\phi}=-\frac{1}{2}\left(\nabla \phi\right)^2 - c_1 e^{c_2 \phi},
\end{eqnarray}
where $ c_1,c_2 >0 $. We do not know yet the relation betweet this Lagrangian and the quintessential black hole solution. However, for the simplest quintessence model, we know that it has no contribution to the total central charge. In addition, the calculation carried out in \cite{Li2010} indicates that in Einstein-Maxwell-Dilaton theory with topological terms in four and five dimensions, the central charge only receives contribution from the gravitational field (see also \cite{Compere2017} for the other proof). So we benefit that the canonical conserved charges from the energy momentum tensor of the quintessence should not be taking into account to calculate the central term. Furthermore, to gain the central charge of the conformal field theory for four-dimensional rotating quintessential black holes, it is sufficient to calculate only the gravitational field contribution. So the contribution of the Maxwell Lagrangian would be put aside. As a proof that we are not supposed to compute the contribution of the quintessence, the resulting entropy from the CFT must be in agreement with the Bekenstein-Hawking one.

\subsection{Central charge}
Because we lose the asymptotically flat space-time in the near-horizon limit, the boundary conditions in spatial infinity are not flat anymore. Thereupon, the choice of boundary conditions of the metric deviation at spatial infinity is needed to be imposed to find the asymptotic symmetry group (ASG). Accordingly, we impose the following boundary conditions
\begin{eqnarray}
h_{\mu \nu} \sim \left(\begin{array}{cccc}
\mathcal{O}(r^2) & \mathcal{O}\left(\frac{1}{r^2}\right) &  \mathcal{O}\left(\frac{1}{r}\right) &  \mathcal{O}(1) \\
 &  \mathcal{O}\left(\frac{1}{r^3}\right) &  \mathcal{O}\left(\frac{1}{r^2}\right) &  \mathcal{O}\left(\frac{1}{r}\right) \\
 &  & \mathcal{O}\left(\frac{1}{r}\right) &  \mathcal{O}\left(\frac{1}{r}\right)\\
 &  &  &  \mathcal{O}(1)\
\end{array} \right), \label{eq:gdeviation}
\end{eqnarray}
in the basis $ (\tau,y,\theta, \phi) $. For all types of $ \omega $, we impose the similar boundary conditions. The more general diffeomorphism symmetry that is consistent with such boundary conditions (\ref{eq:gdeviation}) can be seen in \cite{Guica2009}. Yet it contains one copy of the conformal group of the circle generated by the following Killing field
\begin{eqnarray}
\zeta_n = \varepsilon(\phi)\partial_\phi - y\varepsilon '(\phi)\partial_y ,\label{eq:killingASG}
\end{eqnarray}
where $ \varepsilon(\phi) $ is an arbitrary smooth periodic function of the coordinate $ \phi $ and $ \varepsilon'(\phi)=d\varepsilon/d\phi $. The consistency indicates that the charges associated to the diffeomorphisms have to be finite. Hence we may define $ \varepsilon_n = -e^{-in \phi} $ and $ \zeta_n =\zeta(\varepsilon_n ) $. Because of that, the generator of ASG becomes
\begin{eqnarray}
\zeta_n = -e^{-in \phi}\partial_\phi - inye^{-in \phi}\partial_y ,\label{eq:killingASG1}
\end{eqnarray}
where $ n $ in this case is only an integer. By the Lie bracket, the symmetry generator (\ref{eq:killingASG1}) satisfies the Virasoro algebra
\begin{eqnarray}
i[\zeta_m, \zeta_n]_{LB} = (m-n)\zeta_{m+n},
\end{eqnarray}
without the central term. This algebra only consists of a $ U(1) $, not an $ SL(2,R) $ isometry subgroup.

The diffeomorphim (\ref{eq:killingASG1}) is associated with the charge
\begin{eqnarray}
Q_{\zeta} = \frac{1}{8\pi}\int_{\partial\Sigma} k^{g}_{\zeta}[\mathcal{L}_{\zeta}g;g],
\end{eqnarray}
where
\begin{eqnarray}
k^{g}_{\zeta}[\mathcal{L}_{\zeta}g;g] &=& -\frac{1}{4}\epsilon_{\rho\sigma\mu\nu} \left\{ \zeta ^{\nu} D^{\mu} h - \zeta ^{\nu} D_{\epsilon} h^{\mu \epsilon} + \frac{h}{2} D^{\nu}\zeta^{\mu} \right. \nonumber\\
&&- h^{\nu \epsilon} D_{\epsilon}\zeta^{\mu} + \zeta_{\epsilon}D^{\nu}h^{\mu \epsilon} +\frac{h^{\epsilon \nu}}{2} D^\mu \zeta_\epsilon \nonumber\\
& & \left. +\frac{h^{\epsilon \nu}}{2} D_\epsilon \zeta^\mu \right\} dx^\rho \wedge dx^\sigma , \label{eq:kgrav}\
\end{eqnarray}
and the integral is over the boundary of a spatial slice \cite{Barnich2002,Barnich2008}. Note that the last two terms on (\ref{eq:kgrav}) vanish for an exact Killing vector. The charges from the time-like Killing field and the right-moving part are assumed not to give the contribution as well as the electromagnetic and quintessential fields. Then it follows that the Dirac brackets of the conserved charges are now just the common forms of the Virasoro algebras with a central term such as
\begin{eqnarray}
\{ Q_{\zeta_m},Q_{\zeta_n} \}_{DB} &=& Q_{[\zeta_m ,\zeta_n]} + \frac{1}{8\pi}\int k^{g}_{\zeta_m}\left[\mathcal{L}_{\zeta_n}\bar{g};\bar{g}\right]. \label{eq:charges}\
\end{eqnarray}
Note that $ \bar{g} $ stands for metric background or the extremal rotating quintessential metric. Finally, the central term contains the left-moving central charge $ c_L $ that we need in the Cardy formula which is given as follows
\begin{eqnarray}
\frac{1}{8\pi}\int k^{g}_{\zeta_m}\left[\mathcal{L}_{\zeta_n}\bar{g};\bar{g}\right] = -\frac{c_L}{12}i(m^3+\kappa m)\delta_{m+n,0}, \label{eq:centralchargesterm}\
\end{eqnarray}
where it is set $ \hbar =1 $ and $ \kappa $ is a trivial constant. From the usual Virasoro algebra \cite{Senechal1997} in (\ref{eq:centralchargesterm}), the left-moving central charges for our black hole solutions are
\begin{equation}
c_L^{DD} = 12a\left(M+\frac{\alpha}{2}\right), \label{eq:centalchargeDD}\
\end{equation}
\begin{equation}
c_L^{RD} = 12aM = 12 a \sqrt{a^2+q^2 -n^2 -\alpha}, \label{eq:centalchargeRD}\
\end{equation}
\begin{equation}
c_L^{AE} = \frac{12aM}{(1-\alpha)^2}. \label{eq:centalchargeAE}\
\end{equation}

These left-moving central charges (\ref{eq:centalchargeDD}) and (\ref{eq:centalchargeAE}) are shown in \cite{Sakti2019}. Therein, the central charge is calculated by assuming $ S_{CFT}=S_{BH} $ (see \cite{Skanata2012,Ghezelbash2013,Ghezelbash2014} for the other examples). Herein it is proved that the result of \cite{Sakti2019} matches with the resulting central charge that we compute from the CFT. In general, this central charge will reduce to the one in \cite{Sakti2018} when the quintessential density vanishes.  We do not know what actually the value of $ \alpha $ is because there is no observational evidence of the black holes surrounding by the quintessence until now but we might see the consequence when $ \alpha $ is negative in \cite{Wang2017,Sakti2019}. In this situation if the value of $ \alpha $ is negative, it is needed that the mass $ M $ to be greater than $ \frac{\alpha}{2} $ for unitary CFT. As pointed out in \cite{Siahaan2016} for magnetized Kerr black holes, when $ |B| > 1/M $, the central charge is negative. In fact, a CFT with negative central charge is non-unitary \cite{Senechal1997}.

\subsection{CFT temperature}
In the following, we derive the temperatures from the CFT in the near-horizon region for extremal rotating quintessential black hole. Regarding the Frolov-Thorne vacuum for this black holes, firstly we must recall a quantum scalar field with eigenmodes of the asymptotic energy $ E $ and angular momentum $ J $, which is given by 
\begin{eqnarray}
\tilde{\Phi} = \sum_{E,J,s} \tilde{\phi} _{E,J,s} e^{-i E t + i J \varphi} f_s(r,\theta), \label{eq:quantumfield}\
\end{eqnarray}
for Kerr black hole \cite{Guica2009}. In order to transform this to near-horizon quantities and take the extremal limit, we note that in the near-horizon coordinates (\ref{eq:coortransmatterdom}) we have
\begin{eqnarray}
e^{-i E t + i J \varphi} &=& e^{-i (E-\Omega_H^{ex} J )\tau r_0/\epsilon + i J \phi} \nonumber\\
&=& e^{-in_R \tau + in_L \phi},
\end{eqnarray}
where
\begin{eqnarray}
n_R = (E-\Omega_H^{ex} J)r_0/\epsilon, ~~~n_L = J. \label{eq:nrnl}
\end{eqnarray}
However this is only viable when there is no contribution of the electromagnetic part. On our black hole solution, recall that there are two contributions from the electromagnetic field which are the electric and the magnetic charges, so Eq. (\ref{eq:nrnl}) may be extended to 
\begin{eqnarray}
n_R &=& (E-\Omega_H^{ex} J- \Phi_e^{ex} Q_e- \Phi_g^{ex} Q_g-\Theta_{H}^{ex}\alpha) \frac{r_0}{\epsilon}, \nonumber\\
 n_L &=& J. \label{eq:nRnLnew}
\end{eqnarray}
Hence the corresponding Boltzmann weighting factor is now in the following form
\begin{eqnarray}
&&\exp\left({- \frac{E - \Omega_H^{ex} J - \Phi_e^{ex} Q_e- \Phi_g^{ex} Q_g-\Theta_{H}^{ex}\alpha}{T_H} }\right) \nonumber\\
&&= \exp\left({-\frac{n_R}{T_R}-\frac{n_L}{T_L}-\frac{Q_e}{T_e}-\frac{Q_g}{T_g} -\frac{\alpha}{T_\alpha}}\right)
 \label{eq:Boltzmannweightingfactor}\
\end{eqnarray}
The Boltzmann weighting factor will be a diagonal matrix if we take the trace over the modes inside the horizon. We have to mention that $ T_{L,R} $ is the left-right temperature, $ T_{e,g} $ is the electric-magnetic temperature, and $ T_\alpha $ is the quintessential temperature. We call it as electric-magnetic temperature because it corresponds to the conjugate of electric and magnetic charges. However, in this section we do not need to calculate $ T_{e,g,\alpha} $ because the central charge from the contribution of the Maxwell field is vanishing.

By comparing Eqs. (\ref{eq:nRnLnew}) and (\ref{eq:Boltzmannweightingfactor}), we find the temperatures of the CFT as follows
\begin{eqnarray}
&&T_R = \frac{T_H r_0}{\epsilon}\bigg|_{ex} , ~~~ T_L = - \frac{\partial T_H/\partial r_+}{\partial \Omega_H / \partial r_+}\bigg|_{ex}.\ \label{eq:generalCFTtemperature}
\end{eqnarray}
In the extremal limit, since the right temperature is proportional to the Hawking temperature, it remains zero for all types of $ \omega $
\begin{eqnarray}
T_R^{DD,RD,AE} = 0. \
\end{eqnarray}
So the remaining temperature is the left-moving temperature. They are given by
\begin{equation}
T_L^{DD} = \frac{\left(2M + \alpha \right)^2 +4(a+n)^2}{8\pi a\left(2M + \alpha \right)}, \label{eq:lefttemperatureDD}\
\end{equation}
\begin{equation}
T_L^{RD} = \frac{M^2 +(a+n)^2}{4\pi a M}, \label{eq:lefttemperatureRD}\
\end{equation}
\begin{equation}
T_L^{AE} = \frac{M^2 +(a+n)^2(1-\alpha)^2}{4\pi a M}. \label{eq:lefttemperatureAE}\
\end{equation}
In order to find the positive temperature, we can easily see that for DD type, it must satisfy $ M > \alpha/2 $ when the value of $ \alpha $ is negative. However, there is no observational evidence that shows the value of the quintessential density that surrounds the black holes until now. For RD type, the temperature might be negative if the nominator is negative where $ M^2= a^2+q^2 - n^2 -\alpha$. So the value of $ \alpha $ could be important to determine the sign of the temperature. Fortunately, the temperature remains positive even the quintessential density is negative for AE type because of the square sign.

\subsection{Cardy entropy}
We will use the renowned Cardy formula to match with the Bekensten-Hawking entropy. In the derivation of this formula, the gravitational contribution is turned off \cite{Sakti2018}. Hence this is the reason we might call it as the microscopic entropy. On the other hand, the Bekenstein-Hawking one could be called as the macroscopic entropy. The Cardy formula of the entropy from the CFT is given as follows
\begin{eqnarray}
S = \frac{\pi ^2}{3}(c_L T_L+c_R T_R).\label{eq:cardyentropy}\
\end{eqnarray}
We can compute the microscopic entropy of the extremal rotating quintessence black holes. For every type of $ \omega $, the entropies respectively are
\begin{equation}
S^{DD} = \frac{\pi}{4} \bigg[\left(2M +\alpha\right)^2 + 4(a+n)^2 \bigg] ,\label{eq:entropyDD}\
\end{equation}
\begin{equation}
S^{RD} = \pi \big(2a^2+q^2 +2an-\alpha \big) ,\label{eq:entropyRD}\
\end{equation}
\begin{equation}
S^{AE} = \frac{\pi\big[ M^2 + (a+n)^2(1-\alpha)^2 \big]}{(1-\alpha)^2} .\label{eq:entropyAE}\
\end{equation}
These are the general entropies of the extremal rotating quintessence or KNUTQ black holes with three types of matter domination. It is in agreement with the Bekenstein-Hawking entropy that has been calculated in \cite{Sakti2019}. So it is found that this black hole solution is dual to the CFT. We also prove that the black hole solution from the condition of additivity and linearity for the energy-momentum tensor could still be viable to investigate the correspondence between the gravity and the CFT. If one is interesting with the calculation of the entropy correction, one can check \cite{Sakti2018} where it comes from the assumption of non-zero lowest eigenvalue of the conformal operator.

We can see clearly that when $ \alpha $ goes to zero, it will reduce to the entropy of extremal KNUT black hole \cite{Sakti2018}. If we compare our entropy from the extremal KNUT black hole, we may see that entropy of RD type of black hole is quite similar but the presence of the quintessential density determine the value of the entropy. When it has positive value, the entropy will be smaller than extremal KNUT black hole. Yet the negative value of this density will make the reversed condition happens. 

\section{RNUTQ/CFT Correspondence}

In the previous section, we have derived the entropy of rotating quintessence black hole solution for some types of quintessential equation of state. Something fascinating will occur when we set the rotational spin to be $ 0 $. It should be consistent with the entropy of the extremal Reissner-Nordstr{\"o}m-NUT-Quintessence (RNUTQ) back hole
\begin{eqnarray}
S_{RNUTQ}= \pi\big( r_+^2 + n^2 \big) .\label{eq:entropyRN}\
\end{eqnarray}
But in this manner, the corresponding central charge will reduce to zero because it is proportional to $ a $. The conformal temperature (left-moving temperature) also becomes infinite because it is proportional to $ 1/a $. This is not what we wish to study from the Kerr/CFT correspondence. To solve that problem, we profit from the method used in \cite{Hartman2009,Sakti2018}. They extend the 4-dimensional solution when $ a \rightarrow 0 $ to the 5-dimensional solution as the Kaluza-Klein theory of uplifting dimension. 

In order to do so, we need to introduce the fifth dimensional coordinate or the fibered coordinate of $ S^1 $. This new coordinate is periodic under translation of $ 2\pi R_n $ where $ R_n $ is an integer, so $ z \sim z + 2\pi R_n $. This additional coordinate gives rise to one additional translational Killing field $ \partial_z $. This Killing field enhances the $ SL(2,R)_R\times U(1)_L $ symmetry with another $ U(1) $ symmetry. Besides that, to construct the 5-dimensional extremal metric, we need to map the electromagnetic potential to the geometric part. The detailed derivation will be explained in some next subsections.

\subsection{Near-horizon form of RNUTQ black hole}
Recall that we already define a new fibered coordinate $ z $ and have the 4-dimensional extremal metric $ ds^2 $. After that, the only one that we need is the electromagnetic potential $ \textbf{A} $. However, when we take $ a\rightarrow 0 $, the electromagnetic potential becomes singular. To construct 5-dimensional metric, we are supposed to produce the electromagnetic potential that will not be singular. In order to bring away that singularity, we have to apply the gauge transformation. While we are taking $ a\rightarrow 0 $, we have to apply the following gauged transformation
\begin{eqnarray}
\textbf{A}\rightarrow \textbf{A} - \frac{gn}{a} d\phi. \label{eq:gaugedtrans}
\end{eqnarray}
Finally, the resulting electromagnetic potential for RNUTQ is
\begin{eqnarray}
\textbf{A}&=&\frac{2gnr_+ + e(r_+^2 -n^2)}{V(r_+^2+n^2)}yd\tau  +\frac{(gr_+ - en)\cos \theta}{r_+}d\phi. \label{eq:electromagneticRNUTQ}\
\end{eqnarray}
Now we may produce the 5-dimensional solution using the mapping of the electromagnetic potential (\ref{eq:electromagneticRNUTQ}) to the geometric part as in \cite{Hartman2009,Sakti2018}. Then the following metric is produced
\begin{eqnarray}
ds^2_5&=&\frac{\rho_+^2}{V}\left(-y^2 d\tau^2 +\frac{dy^2}{y^2}+Vd\theta ^2 + V\sin^2\theta d\phi^2 \right) \nonumber\\
& &+  \left(dz + \frac{2gnr_+ + e(r_+^2 -n^2)}{V(r_+^2+n^2)}yd\tau \right. \nonumber\\
&&\left. +\frac{(gr_+ - en)\cos \theta}{r_+}d\phi \right)^2,
\end{eqnarray}
where $\rho_+^2 = r_+^2 + n^2 $. Recall that for dust domination $ \omega = 0 $, we have $ r_+ = ( M +\alpha/2) $ and $ V=1 $. For radiation domination $ (\omega=1/3) $, we have $ r_+ = M  $ and $ V=1 $ which is similarly with extremal RNUT black hole \cite{Sakti2018}. The last one is the accelerating expansion type $ (\omega=-1/3) $ which have $ r_+ = \frac{M}{1-\alpha} $ and $ V=1-\alpha $. The presence of fibered coordinate will enhance the solution to have global $ U(1)_{gauge} $ symmetry.

\subsection{Central charge}
In the following, we will compute the central charge of the 5-dimensional solution. Since this solution is the extension of the 4-dimensional solution of KNUTQ black hole, we could apply the similar way to find the central charge. As the 4-dimensional solution, the central charge comes from the gravitational part only, so we will neglect the contribution of the electromagnetic potential and the quintessential field.  This benefits us again because there is no explicit Lagrangian form of the quintessence. 

Looking for the non-trivial diffeomorphisms, we need to set some consistent boundary conditions of the metric deviation as in 4D solution. So we will adopt the same boundary conditions such in \cite{Hartman2009,Sakti2018}
\begin{eqnarray}
\small
h_{\mu \nu} \sim \left(\begin{array}{ccccc}
\mathcal{O}(y^2) & \mathcal{O}\left(\frac{1}{y^2}\right) &  \mathcal{O}\left(\frac{1}{y}\right) & \mathcal{O}(y) &  \mathcal{O}(1) \\
 &  \mathcal{O}\left(\frac{1}{y^3}\right) &  \mathcal{O}\left(\frac{1}{y^2}\right) & \mathcal{O}\left(\frac{1}{y}\right) &  \mathcal{O}\left(\frac{1}{y}\right) \\
 &  & \mathcal{O}\left(\frac{1}{y}\right) & \mathcal{O}(1) &  \mathcal{O}\left(\frac{1}{y}\right)\\
 &  &  & \mathcal{O}\left(\frac{1}{y}\right) & \mathcal{O}(1)\\
 &  &  &  & \mathcal{O}(1)\
\end{array} \right), \label{eq:gdeviation2}
\end{eqnarray}
in the basis $ (\tau,y,\theta,\phi, z) $. Performing those boundary conditions, then we find the most general diffeomorphisms which are given by
\begin{eqnarray}
\zeta &=& \left\{b_\tau + \mathcal{O}\left(y^{-3}\right) \right\}\partial _\tau + \left\{-y\epsilon '(z) + \mathcal{O}(1) \right\}\partial _y  \nonumber\\
&&+ \mathcal{O}\left(y^{-1}\right)\partial _\theta + \left\{b_{\phi} + \mathcal{O}\left(y^{-2}\right) \right\}\partial _\phi \nonumber\\
&& + \left\{ \epsilon(z) + \mathcal{O}\left(y^{-2}\right) \right\}\partial _z,\
\end{eqnarray}
where $ \epsilon (z) = -e^{-inz} $ and the prime $ (') $ denotes the derivative respect to $ z $. Here $ b_\tau, b_\phi $ are arbitrary constants. Therefore the asymptotic symmetry group contains one copy of the conformal group of the circle generated by
\begin{eqnarray}
\zeta_z = \epsilon(z) \partial_z - y \epsilon '(z)\partial _y ,\
\end{eqnarray}
but do not allow $ \zeta_\epsilon $ such the case of 4-dimensional KNUTQ solution. Moreover, to compute the central charge, we may follow the same steps such for KNUTQ solution but for five dimensional gravity \cite{Ghodsi2010,Sakti2018}.  We apply
\begin{eqnarray}
k^{g}_{\zeta}[h;g] &=& -\frac{1}{2}\frac{1}{3!}\epsilon_{\rho\sigma \gamma \mu\nu} \left\{ \zeta ^{\nu} D^{\mu} h - \zeta ^{\nu} D_{\lambda} h^{\mu \lambda} + \frac{h}{2} D^{\nu}\zeta^{\mu}  \right. \nonumber\\
&& - h^{\nu \lambda} D_{\lambda}\zeta^{\mu} + \zeta_{\lambda}D^{\nu}h^{\mu \lambda} +\frac{h^{\lambda \nu}}{2}D^\mu \zeta_\lambda  \nonumber\\
& & + \frac{h^{\lambda \nu}}{2} D_\lambda \zeta^\mu \bigg\} dx^\rho \wedge dx^\sigma \wedge dx^\gamma , \label{eq:kgrav2} \
\end{eqnarray}
where the last two terms vanish for the exact Killing vector. In the end, we will find
\begin{equation}
c^{DD}_z = 6\left[ gn(2M+\alpha) +e \left\{\left(M +\frac{\alpha}{2}\right)^2 -n^2 \right\} \right], \label{eq:centralchargeRNUTQ1}\
\end{equation}
\begin{equation}
c^{RD}_z = 6\left[ 2gnM +e (M^2 -n^2) \right], \label{eq:centralchargeRNUTQ2}\
\end{equation}
\begin{equation}
c^{AE}_z = \frac{6}{1-\alpha}\left[ \frac{2gnM}{1-\alpha} +e\left(\frac{M^2}{(1-\alpha)^2} -n^2 \right) \right]. \label{eq:centralchargeRNUTQ3}
\end{equation}
These all central charges correspond to $ \zeta_z $ and represent all type quintessential equation of state which we choose.

\subsection{Temperature}
Since the spin is assumed to vanish, the corresponding thermodynamic potential for the angular momentum also vanishes. The temperatures conjugate to electric and magnetic charges are the only temperatures which are left now. Recall the first law of black hole thermodynamics, in this situation we have
\begin{eqnarray}
dS = \frac{dQ_e}{T_e}+\frac{dQ_g}{T_g}+\frac{d\alpha}{T_\alpha}.
\end{eqnarray}
As the second dual CFT for extremal RN-AdS \cite{Hartman2009} and RNUT solutions \cite{Sakti2018}, the magnetic charge is held fixed. Because there exists a quintessential density, the vacuum is not a pure state as the extremal Kerr-Newman-AdS and extremal Kerr-Newman-NUT black holes \cite{Hartman2009,Sakti2018}. The remaining temperatures are given by
\begin{equation}
T^{DD}_e =\frac{\left(2M+\alpha\right)^2 +4n^2}{8\pi \left[ gn(2M+\alpha)+e \left\{\left(M +\frac{\alpha}{2}\right)^2 -n^2 \right\} \right]}, \label{eq:temperatureofRNUTQ1}\
\end{equation}
\begin{equation}
T^{RD}_e =\frac{M^2 +n^2}{2\pi \left[ 2gnM+e (M^2 -n^2) \right]},\label{eq:temperatureofRNUTQ2}\
\end{equation}
\begin{equation}
T^{AE}_e =\frac{M^2 +n^2(1-\alpha)^2}{2\pi(1-\alpha) \left[ \frac{2gnM}{1-\alpha} +e \left(\frac{M^2}{(1-\alpha)^2} -n^2\right) \right]}. \label{eq:temperatureofRNUTQ3}\
\end{equation}
These temperatures are different with left-moving one because it comes from the conjugate electric charge. For the case of 4-dimensional KNUTQ black hole, we have mentioned that there are $ T_{e},T_{g},T_\alpha $. Since in 5-dimensional solution $ T_{L} $ is singular when $ a \rightarrow 0 $, we need to apply another temperature to Cardy formula which are $ T_{e},T_{g},T_\alpha $.

\subsection{Entropy}
Having the conjugate temperature of the electric charge and the central charge from 5-dimensional solution, we are able to compute the microscopic entropy of the RNUTQ black hole by performing the famous Cardy formula. For this 5-dimensional solution, we possess
\begin{eqnarray}
S_{CFT} = \frac{\pi^2}{3}c_z T_e. \label{eq:Cardy5dimensional}\
\end{eqnarray}
Using the central charges (\ref{eq:centralchargeRNUTQ1})-(\ref{eq:centralchargeRNUTQ3}) and CFT temperatures (\ref{eq:temperatureofRNUTQ1})-(\ref{eq:temperatureofRNUTQ3}) to the formula (\ref{eq:Cardy5dimensional}), the entropies of the extremal RNUTQ for every type of black hole are then 
\begin{equation}
S^{DD} = \frac{\pi}{4} \bigg[\left(2M +\alpha\right)^2 + 4n^2 \bigg] ,\label{eq:entropyRNDD}\
\end{equation}
\begin{equation}
S^{RD} = \pi \big(q^2-\alpha \big) ,\label{eq:entropyRNRD}\
\end{equation}
\begin{equation}
S^{AE} = \frac{\pi\big[ M^2 + n^2(1-\alpha)^2 \big]}{(1-\alpha)^2} .\label{eq:entropyRNAE}\
\end{equation}
Therefore the entropy from the CFT for this 5D solution also matches the Bekenstein-Hawking entropy. To obtain the entropy of extremal RNUT black hole \cite{Sakti2018}, we may set $ \alpha = 0 $. In addition to that, taking zero NUT charge will reduce the entropy to the entropy of extremal Reissner-Nordstr{\"o}m black hole \cite{Ghodsi2010}. This entropy for RNUTQ black holes shows that this black hole solution is dual to the CFT.

\section{Summary}
In this article, we have investigated the Kerr/CFT correspondence on extremal rotating quintessential black hole. We analyze the near-horizon extremal geometry of this quintessential black holes with specific values of quintessential EoS. We have chosen the EoS parameter ($ \omega = 0, 1/3,-1/3 $) to prevent from getting the solution with four horizons. After getting the near-horizon form, we calculate the central charge. A problem appears because there is no specific Lagrangian form or quintessence that could be used to derive canonical conserved charge to compute the central charge. However, we know that the simplest quintessence model could be described with a real scalar field with the canonical kinetic term and an exponential potential. Furthermore, we benefit of some recent calculation that have shown that the contribution of non-gravitational fields is absent. So we do not need to compute the contribution of the quintessence by assuming it will vanish. Then we employ the ASG to calculate the central charges coming from the central term of the Dirac bracket of the conserved charges. It is found that the central charge contains quintessential density parameter. For DD type, the mass of the black hole should be greater than $ \alpha/2 $ to get non-negative central charge if the value of quintessential density is negative. We also confirm the result of the central charge which is obtained in another article where the dual CFT is not applied yet. Then adopting the Frolov-Thorne vacuum, the CFT temperatures are derived in order to compute the microscopic entropy. Again for DD type, the similar assumption should be applied as for the central charge in order to obtain positive temperature. Then applying Cardy formula, it is obtained that the entropy from the CFT matches the Bekenstein-Hawking entropy for the extremal rotating quintessential black holes. So obviously we see that the extremal KNUTQ black hole is dual with the CFT even it has no exact Lagrangian form of the quintessence.

We also extend the solution to 5-dimensional Reissner-Nordstr{\"o}m-NUT-Quintessence black holes. The reason of doing this is because there is a problem when we take $ a\rightarrow 0 $. We can still get the exact entropy although it makes the central charge is going to zero and the left-moving temperature is becoming infinite. So we employ the similar way as for the Kerr-Newman-AdS and Kerr-Newman-NUT black holes to find RN-AdS and RNUT solutions to solve this problem. We map the electromagnetic potential as the geometric part and add a new fifth coordinate to construct this extremal solution. After imposing some boundary conditions for ASG, we find the central charge related to the fifth coordinate for all types of EoS parameter. We also find the temperatures which are the conjugate of the electric charge only because we hold the magnetic charge fixed. Finally, performing Cardy formula for all types, we obtain the entropy of extremal RNUTQ black holes. One can see that the extremal RNUTQ black hole solution is also dual with the CFT. Then it would be fascinating to study the similar correspondence of this quintessential black hole solutions with the exact Lagrangian form to relate the result which we find in this paper.

\begin{acknowledgement}
\textbf{Acknowledgement} We gratefully acknowledge the support from Ministry of Research, Technology, and Higher Education of the Republic of Indonesia and PMDSU Research Grant. M.F.A.R.S. also wants to express many thanks all members of Theoretical Physics Laboratory, Institut Teknologi Bandung for the valuable support.
\end{acknowledgement}

%
%
%
%

\end{document}